# Pressure effect on superconductivity in FeSe$_{0.5}$Te$_{0.5}$

**Sergii I. Shylin**[*,1,2], **Vadim Ksenofontov**[*,1], **Pavel G. Naumov**[3], **Sergey A. Medvedev**[3], **Vladimir Tsurkan**[4,5], **Joachim Deisenhofer**[4], **Alois Loidl**[4], **Leslie M. Schoop**[6], **Taras Palasyuk**[7], **Gerhard Wortmann**[8] and **Claudia Felser**[3]

[1] Institute of Inorganic and Analytical Chemistry, Johannes Gutenberg University Mainz, Staudingerweg 9, D-55099 Mainz, Germany
[2] Department of Chemistry, Taras Shevchenko National University of Kyiv, Volodymyrska 64/13, 01601 Kyiv, Ukraine
[3] Max Planck Institute for Chemical Physics of Solids, D-01187 Dresden, Germany
[4] Experimental Physics V, University of Augsburg, D-86159 Augsburg, Germany
[5] Institute of Applied Physics, Academy of Sciences of Moldova, MD 2028, Chisinau, Moldova
[6] Max Planck Institute for Solid State Research, D-70569 Stuttgart, Germany
[7] Institute of Physical Chemistry, Polish Academy of Sciences, Kasprzaka 44/52, 01-224 Warsaw, Poland
[8] Department of Physics, University of Paderborn, D-33095 Paderborn, Germany

Due to the simple layered structure, isostructural FeSe and FeSe$_{0.5}$Te$_{0.5}$ are clue compounds for understanding the principal mechanisms of superconductivity in the family of Fe-based superconductors. High-pressure magnetic, structural and Mössbauer studies have been performed on single-crystalline samples of superconducting FeSe$_{0.5}$Te$_{0.5}$ with $T_c$ = 13.5 K. Susceptibility data have revealed a strong increase of $T_c$ up to 19.5 K for pressures up to 1.3 GPa, followed by a plateau in the $T_c(p)$ dependence up to 5.0 GPa. Further pressure increase leads to a disappearance of the superconducting state around 7.0 GPa. X-ray diffraction and Mössbauer studies explain this fact by a tetragonal-to-hexagonal structural phase transition. Mössbauer parameters of the non-superconducting high-pressure phase indicate less covalency of Fe–Se bonds. Based on structural and susceptibility data we conclude about a common character of $T_c(p)$ diagrams for both FeSe and FeSe$_{0.5}$Te$_{0.5}$ superconductors.

**1 Introduction** Discovery of superconducting compounds A$_x$Fe$_{2-y}$Se$_2$ with intercalated alkali elements (A = K, Rb, Cs) has extended superconducting transition temperature above 30 K and confirmed an essential research potential of FeSe-based systems [1]. The nearly stoichiometric FeSe becomes superconducting below 8 K at ambient pressure but the transition temperature can be enhanced up to 37 K by application of external pressure of ca. 9 GPa [2, 3]. Above this pressure, FeSe transforms to a hexagonal close packed NiAs-type structure that exhibits semiconducting behaviour. Variety of experiments demonstrates ambiguous relation between the crystal structures of FeSe-based compounds and their superconducting properties. For instance, the superconducting transition in Fe$_{1.01}$Se occurs in an orthorhombic phase which appears after a subtle structural transition from the tetragonal phase at ca. 90 K [4]. Below this temperature Fe$_{1.01}$Se exhibits nematic ordering without long-range magnetism, which competes with emerging superconductivity [5]. In contrast, the non-superconducting composition Fe$_{1.03}$Se remains always in the tetragonal phase. Pressure application at room temperature converts Fe$_{1.01}$Se to the non-metallic NiAs-type polymorph around 9 GPa, which remains stable at low temperatures [2]. The substitution of Se by Te in FeSe$_{1-x}$Te$_x$ increases the superconducting transition temperature up to 14 K for x = 0.5 [6, 7]. The low-temperature structural response to pressure of orthorhombic FeSe$_{0.55}$Te$_{0.42}$ is different than in FeSe as evidenced by the observation of a monoclinic structure at 2.3 GPa, the same pressure, where $T_c$ = 23.3 K is found to reach its maximum [8]. Further pressure increase leads to a monotonic decrease of $T_c$ in the monoclinic phase that exists up to a pressure of 11.9 GPa, at which superconductivity is completely suppressed. According to resistivity data, above this pressure FeSe$_{0.55}$Te$_{0.42}$ remains metallic [8]. Formally FeSe, FeSe$_{0.55}$Te$_{0.42}$, FeSe$_{0.5}$Te$_{0.5}$, as well as Cu-substituted FeSe, reveal similar dome-shape curves in their $T_c(p)$ diagrams limiting the range of superconductivity [2, 8, 9]. In the previous study of FeSe$_{0.5}$Te$_{0.5}$ it was found that $T_c$ increases rapidly from 13.5 K to 26.2 K upon applying pressures up to 2 GPa. Above 2 GPa, $T_c$ decreases linearly and a non-superconducting metallic phase is observed at $p$ = 14 GPa [10]. The authors point out that the same relationship between normalized $T_c$ and pressure in both FeSe$_{0.5}$Te$_{0.5}$ and FeSe presumes universal pressure dependence in these systems, but suggest that the phase transition from the tetragonal to the hexagonal modification observed in FeSe does not occur in FeSe$_{0.5}$Te$_{0.5}$. Considering an interest to superconducting FeSe$_{0.5}$Te$_{0.5}$ compound and taking into account a scope of experimental data available [11-16], a complete $T_c(p)$ diagram reflecting the interrelation of structural, electronic and superconducting properties would be demanded. Additional experimental information is also necessary to clarify the unusual rapid growth, subsequent stagnation and disappearance of $T_c$ in FeSe$_{1-x}$Te$_x$ under pressure.

It has been shown earlier that no significant pressure variations that can be responsible for the initial rapid increase of $T_c$ with pressure occur in the phonon spectrum of FeSe [17]. In this context, the suggestion that the strong enhancement of $T_c$ under pressure in FeSe$_{0.5}$Te$_{0.5}$ is mainly due to an increase of density of electronic states [18], requires further consideration. Until now the superconducting properties of powdered FeSe$_{0.5}$Te$_{0.5}$ samples were investigated only by resistivity measurement under pressure [8]. The disadvantage of this method is the influence of particle boundaries, distribution of particles size and percolation effects on the formation of the $T_c(p)$ curve. Magnetic susceptibility measurements of single crystals under pressure should be more appropriate to obtain a reliable $T_c(p)$ dependence. We also present here results of high-pressure structural and $^{57}$Fe Mössbauer studies of superconducting single crystalline FeSe$_{0.5}$Te$_{0.5}$.

**2 Experimental details** Single crystals of FeSe$_{0.5}$Te$_{0.5}$ were grown by Bridgman method. Details of the preparation and sample characterization were described elsewhere [16]. The quality of the grown samples was confirmed by the X-ray diffraction and magnetic measurements.

High-pressure X-ray diffraction experiments were performed at room temperature on the BL12B2 beamline at SPring-8 synchrotron facility, Japan. For X-ray diffraction the grained sample of FeSe$_{0.5}$Te$_{0.5}$ was loaded in a diamond anvil cell with at culets of diameter 450 μm and a tungsten gasket with sample chamber of diameter 150 μm. Silicon oil was used as a pressure transmitting medium. X-ray beam was collimated to 100 μm. As detector, the ADSC Quantum 210r image plate reader was set up perpendicular to the beam path. Typical accumulation time of diffraction pattern was 2 min. Cerium dioxide was used as an external standard to determine the beam center, sample-to-detector distance, exact wavelength ($\lambda$ = 0.56289 nm) and tilting angle of the image plate. Collected full-circle powder patterns were processed with FIT2D software.

Magnetic susceptibility measurements under pressure were performed using a SiC-anvil high-pressure cell made from a non-magnetic hardened Cu-Ti alloy equipped with SiC anvils. The diameter of the flat working surface of the SiC-anvil was 0.8 mm, and the diameter of the hole in the gasket was 0.3 mm. The cell allows quasihydrostatic pressures up to 12 GPa [19]. The hole was filled with the crystalline FeSe$_{0.5}$Te$_{0.5}$ sample and Daphne oil as a pressure transmitting medium. Pressure was measured by the Ruby scale from a small chips scattered across the sample. The pressure inhomogeneity was estimated 0.5 GPa across the sample. $T_c$ was determined from the onset of the superconducting transition curve, i.e. from the intersection of two extrapolated straight lines drawn through the curve in the normal state and the one drawn through the steepest part of the curve in the superconducting state.

$^{57}$Fe-Mössbauer spectra were recorded using a constant-acceleration spectrometer and a $^{57}$Co(Rh) Mössbauer source with an active spot diameter of 0.5 mm. The spectrometer was equipped with a helium bath cryostat operating in 5 – 300 K temperature range. The Mössbauer absorber of single crystalline FeSe$_{0.5}$Te$_{0.5}$ was prepared by the so-called scotch-tape technique [20], e.g. extracting thin sample layers from the crystal by a scotch tape and inserting 4 sample layers fixed on the tape in the absorber holder. For the Mössbauer measurements under pressure, grained $^{57}$Fe-enriched (20%) FeSe$_{0.5}$Te$_{0.5}$ sample was loaded in a diamond anvil pressure cell with silicon oil as the pressure transmitting medium enabling quasihydrostatic pressure. The isomer shift values were quoted relative to α-Fe at 295 K.

**3 Results**
**3.1 Mössbauer spectroscopy characterization**
Preparation of single crystal sample of FeSe$_{0.5}$Te$_{0.5}$ by scotch-tape technique results in a highly textured FeSe$_{0.5}$Te$_{0.5}$ absorber with the *c* axis of the tetragonal structure oriented preferentially parallel to the transmitting gamma rays. The Mössbauer spectra shown in Fig. 1 exhibit an asymmetric quadrupole doublet with an intensity ratio I$_-$ / I$_+$ of about 2.1. This value allows to derive a negative sign of $V_{zz}$, the electric field gradient, and of the quad-rupole splitting $\Delta E_Q = e^2QV_{zz}/2$. The derived hyperfine parameters at room temperature, isomer shift $\delta = 0.463(1)$ mm s$^{-1}$ and $|\Delta E_Q| = 0.283(2)$ mm s$^{-1}$, are close to those for superconducting FeSe [21, 22] and indicate a low-spin state of divalent iron in tetrahedral chalcogen environment. The asymmetry of the quadrupole doublet arises solely from the highly textured absorber, which is strikingly demonstrated by measuring the absorber tilted by 54.7°, the so-called magic angle, with respect to the gamma rays. The symmetric quadrupole doublet (Fig. 1b) exhibits an intensity ratio of 1:1, indicating absence of any impurity phases or variety of Fe sites.

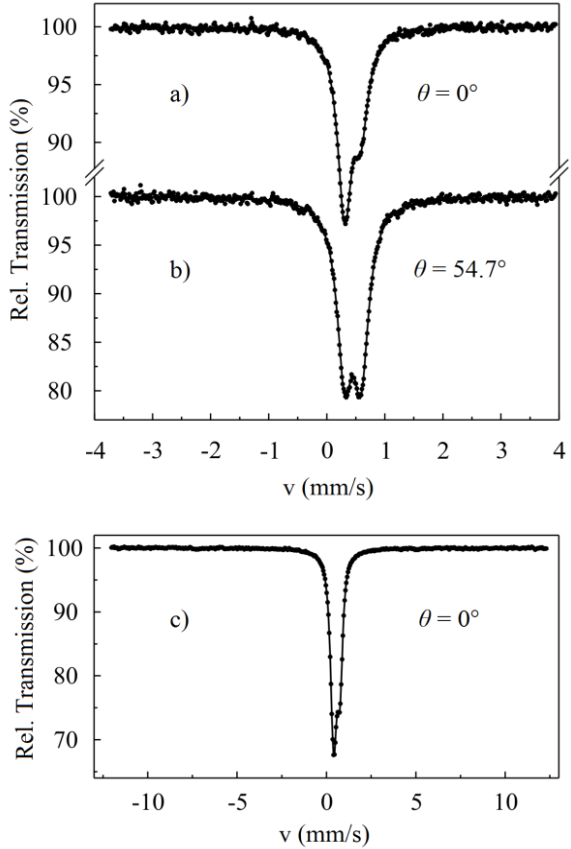

**Figure 1** $^{57}$Fe-Mössbauer spectra of single crystalline FeSe$_{0.5}$Te$_{0.5}$ acquired in transmission geometry with a wave vector of γ-rays perpendicular to the sample plane at 293 K (a) and 5 K (c); spectrum measured at magic angle $\theta = 54.7°$ at 293 K (b).

The temperature dependent spectra were measured between 5 K to 295 K in the large velocity range shown in Fig. 1c and fitted with a routine using a Voigt profile (convolution of Lorentzian and Gaussian profiles) to obtain reliable values of the hyperfine parameters for the thick absorber. The derived values for the normalized spectral area, the isomer shift, $\delta$, and the quadrupole splitting, $|\Delta E_Q|$, are plotted in Fig. 2. Experimental results do not show any feature in the vicinity of $T_c$ and do not support any scenario of superconductivity based on anomalous softening of a phonon spectrum. The gradual decrease of $\delta$ with increasing temperature is caused by the second-order Doppler shift only and can be described by a Debye model [23]. Temperature dependence of $|\Delta E_Q|$ is also typical for iron chalcogenides and can be plausibly fitted using the simplified model for tetragonal distortion in an axial electric field [21].

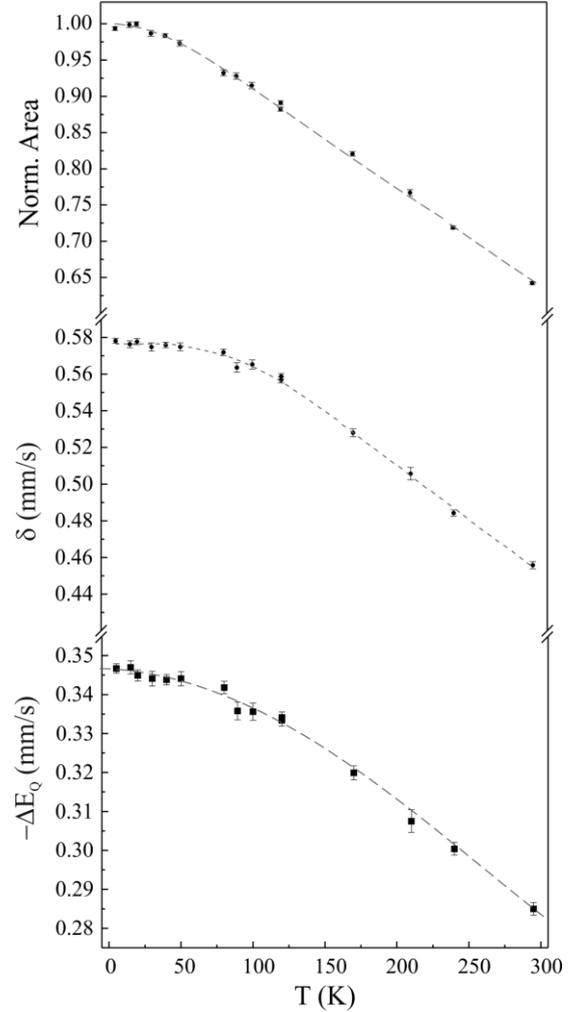

**Figure 2** Temperature dependences of the normalized spectral area, $\delta$ and $\Delta E_Q$ for FeSe$_{0.5}$Te$_{0.5}$. Fitting details are described in the text.

**3.2 Magnetic studies** Fig. 3 shows the temperature dependence of the normalized magnetization of FeSe$_{0.5}$Te$_{0.5}$ at various pressures in a magnetic field of 20 Oe after zero-field cooling (ZFC). A fast increase of $T_c$ with increasing pressure is observed up to 1.3 GPa. In the pressure range of 1.3 GPa to 4.8 GPa, $T_c$ attains a value of approximately 20 K and remains essentially pressure-independent, giving rise to a distinctive plateau in the $T_c(p)$ dependence. Above 5 GPa, $T_c$ starts to decrease reaching 17.2 K at 6.4 GPa. At 7.6 GPa no superconductivity is observed above 2.0 K.

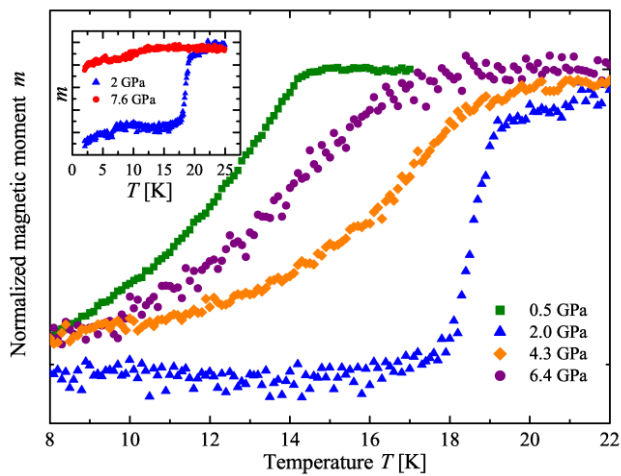

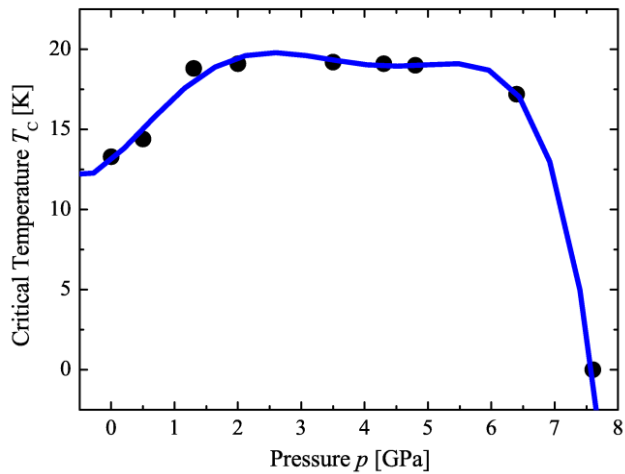

**Figure 3** Selected magnetization curves of FeSe$_{0.5}$Te$_{0.5}$ measured at different pressures in a 20 Oe magnetic field (ZFC). Data are normalized to a maximal signal value for every pressure. At pressures above 7.6 GPa superconductivity disappears (inset).

**Figure 4** Pressure dependence of $T_c$ (onset) for FeSe$_{0.5}$Te$_{0.5}$ obtained by magnetization measurements. Solid line is a guide to the eye.

Comparison of the present $T_c(p)$ curve (Fig. 4) to the corresponding experimental curve obtained from resistivity measurements [10] shows that in both cases a fast increase of $T_c$ with pressure up to 2 GPa is observed. However, this increase is more pronounced in the work of Horigane et al. [10], where maximal value of $T_{onset}$ = 26.2 K is reported. The maximal value of $T_c$ = 19.5 K in the present magnetization measurements is substantially lower, and a $T_c(p)$ dependence exhibits a much more flat and clear plateau. A second difference in the two studies of FeSe$_{0.5}$Te$_{0.5}$ is the fast disappearance of $T_c$ above 7.6 GPa in the magnetization measurements, while from the resistivity data the loss of superconductivity is extrapolated to occur at 9.5 GPa or 12 GPa from the $T_{offset}$ and $T_{onset}$ data respectively. However, due to the small sample volume which is possible to load into the high pressure cell used for magnetization measurements it is not possible to observe small superconducting fractions of sample. The disappearance of superconductivity in our magnetization measurements looks like the disappearance of bulk superconductivity in a single crystal. Other resistivity data [8, 24] are similar to those of [10], showing a slow decrease of $T_c(p)$ dependence up to 12 GPa.

**3.3 XRD studies** The synchrotron X-ray powder diffraction pattern recorded at the lowest experimental pressure of 0.4 GPa and at room temperature could be indexed with a tetragonal *P4/nmm* lattice (anti-PbO type) with the lattice parameters a=3.788 Å and c=5.884 Å that is in good agreement with ambient pressure structural data [25]. The diffraction patterns indicate that tetragonal anti-PbO type structure of FeSe$_{0.5}$Te$_{0.5}$ remains stable at pressures below 6.5 GPa. At higher pressures the onset of the structural phase transition occurs (Fig. 5).

The diffraction patters of the high-pressure phase collected at 19.0 GPa can be assigned to a NiAs structure (*P6$_3$/mmc*), similar to the high-pressure phase in FeSe [2]. Similarly to FeSe, the phase transition is associated with significant reduction (15%) of the unit cell volume. Thus, the structural response of FeSe$_{0.5}$Te$_{0.5}$ on compression at room temperature observed here is identical to that of the FeSe but it differs from the phase sequence observed at low-temperature compression of Fe$_{1.03}$Se$_{0.57}$Te$_{0.43}$ [8] where pressure above ca. 3.0 GPa caused a discontinues transformation of the low-temperature orthorhombic phase into the monoclinic phase which remains stable up to 14.0 GPa.

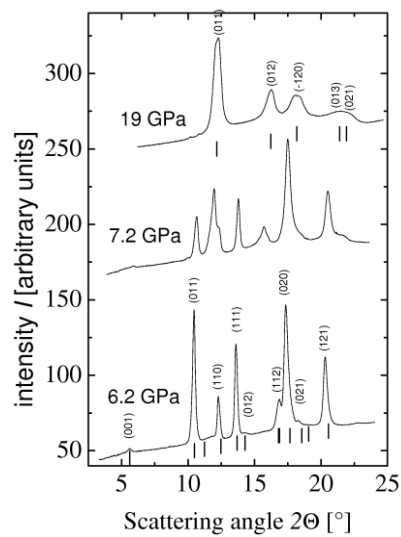

**Figure 5** Room temperature synchrotron X-ray powder diffraction patterns of FeSe$_{0.5}$Te$_{0.5}$ recorded at 6.2 GPa (tetragonal phase), 7.2 GPa (coexisting tetragonal and hexagonal phases) and 19.0 GPa (pure hexagonal high-pressure phase).

**3.4 Mössbauer studies under pressure** Selected Mössbauer spectra of $FeSe_{0.5}Te_{0.5}$ at pressures up to 15.7 GPa are shown in Fig. 6. Below 5.5 GPa, a single quadrupole doublet is observed, which corresponds to the tetragonal phase. Similarly to FeSe [2], at pressure of 8.3 GPa another quadrupole doublet with $\delta = 0.51(2)$ mm s$^{-1}$ and $|\Delta E_Q| = 0.42(3)$ mm s$^{-1}$ appears, which should be assigned to the high-pressure hexagonal phase. The coexistence of both phases is observed up to 10.5 GPa, and at higher pressures spectra can be fitted using single doublet that is consistent with the structural measurements under pressure.

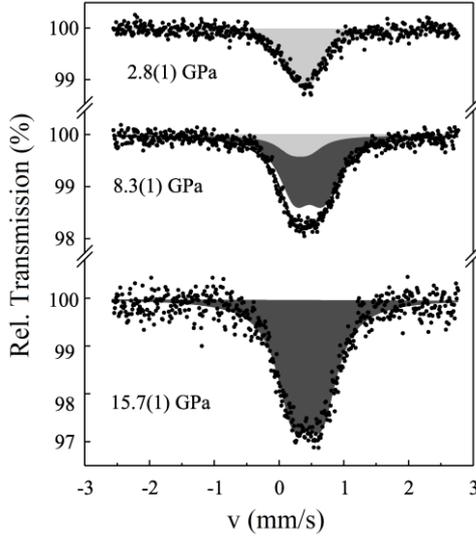

**Figure 6** $^{57}$Fe-Mössbauer spectra of $FeSe_{0.5}Te_{0.5}$ at 295 K measured at 2.8 GPa, 8.3 GPa and 15.7 GPa. Light-gray shading corresponds to the tetragonal phase. The dark-gray shading indicates the doublet of the high-pressure hexagonal phase.

Fig. 7 shows the pressure dependence of the isomer shift and quadrupole splitting of iron in the tetragonal and hexagonal phases of $FeSe_{0.5}Te_{0.5}$, measured at room temperature. The rates of the hyperfine parameters changes in tetragonal phase are $(\partial\delta/\partial p) = -0.011(3)$ and $(\partial\Delta E_Q/\partial p) = 0.002(1)$ mm s$^{-1}$ GPa$^{-1}$. These numbers are close to the corresponding values for FeSe $(\partial\delta/\partial p) = -0.015(3)$ and $(\partial\Delta E_Q/\partial p) = 0.001(4)$ mm s$^{-1}$ GPa$^{-1}$. For the hexagonal phase, pressure dependence of $\delta$ and $\Delta E_Q$ can be formally described in linear approximation by $(\partial\delta/\partial p) = -0.004(7)$ and $(\partial\Delta E_Q/\partial p) = 0.002(5)$ mm s$^{-1}$ GPa$^{-1}$. The hyperfine parameters in the hexagonal phase correspond to divalent iron with less covalency comparing to the tetragonal phase similarly to FeSe [2] (Table 1).

As well as in superconducting FeSe, the distortion of the local surrounding of Fe atoms in $FeSe_{0.5}Te_{0.5}$ is higher in the high-pressure phase. A decrease of pressure exhibits a hysteresis with a width of 3.3 GPa which suggests a first-order type pressure induced structural transition. When pressure is released, the original tetragonal phase is restored without any indication of the high-pressure phase. The relatively small change of the hyperfine parameters in both tetragonal and hexagonal phases with pressure indicates a modest variation of the local surroundings of the Fe ions.

**Table 1** Comparison of the Mössbauer parameters for $FeSe_{0.5}Te_{0.5}$ and FeSe [2, 21] in the tetragonal (t) and hexagonal (h) phases at room temperature.

| Parameter | $FeSe_{0.5}Te_{0.5}$ | FeSe |
|---|---|---|
| $\delta_t$ (1 bar) [mm s$^{-1}$] | 0.463(1) | 0.450(2) |
| $|\Delta E_Q|_t$ (1 bar) [mm s$^{-1}$] | 0.283(2) | 0.245(3) |
| $(\partial\delta_t/\partial p)$ [mm s$^{-1}$ GPa$^{-1}$] | $-0.011(3)$ | $-0.015(3)$ |
| $(\partial\Delta E_{Qt}/\partial p)$ [mm s$^{-1}$ GPa$^{-1}$] | 0.002(1) | 0.001(4) |
| $\delta_h$ (8.3 GPa) [mm s$^{-1}$] | 0.51(3) | 0.58(5) |
| $|\Delta E_Q|_h$ (8.2 GPa) [mm s$^{-1}$] | 0.42(3) | 0.52(7) |
| $(\partial\delta_h/\partial p)$ [mm s$^{-1}$ GPa$^{-1}$] | $-0.004(7)$ | $-0.005(9)$ |
| $(\partial\Delta E_{Qh}/\partial p)$ [mm s$^{-1}$ GPa$^{-1}$] | 0.002(5) | 0.001(1) |

A decrease in the isomer shift for both sites is observed, which corresponds to an increase in the $s$-electron density at the Fe nuclei under pressure. There are several mechanisms by which the core electron contribution $|\psi(0)|^2$ can be altered by pressure [26]. The $s$-like conduction electrons behave approximately like a free electron gas so that changes in core electron contribution should scale nearly inversely with a volume: $|\psi(0)|^2 \sim 1/V$. Summarizing Mössbauer data, the transformations seen in the spectra can be associated with a first order structural phase transition between the low-pressure tetragonal and the high-pressure hexagonal modifications of $FeSe_{0.5}Te_{0.5}$. Since superconductivity disappears above 7.6 GPa, the high-pressure phase is also stable at low temperatures, which explains the loss of superconductivity.

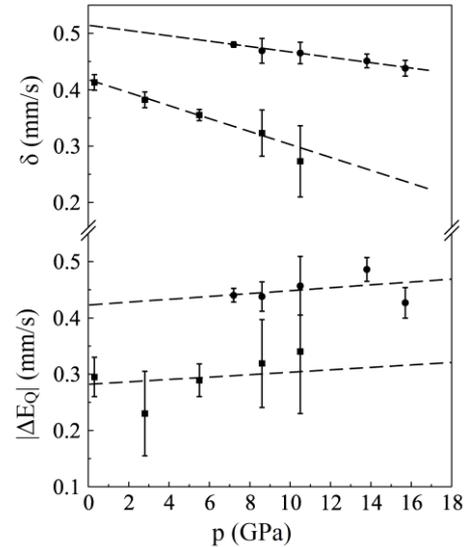

**Figure 7** Pressure dependence of the isomer shifts, $\delta$, and quadrupole splitting, $|\Delta E_Q|$, of Fe in tetragonal (closed squares) and hexagonal (closed cycles) sites of $FeSe_{0.5}Te_{0.5}$ at room temperature obtained at monotonously ascending pressure. The data at 7.2 GPa, which show existence of hexagonal phase only, have been measured after releasing pressure from 15.7 GPa.

**4 Discussion** Pressure dependence of the tetragonal phase fraction obtained from Mössbauer measurements at room temperature for FeSe (closed circles) and FeSe$_{0.5}$Te$_{0.5}$ (open circles) together with the normalized transition temperature $T_c/T_{c,max}$ for FeSe$_{0.5}$Te$_{0.5}$ derived from magnetization measurements are presented in Fig. 8. An increase of the hexagonal phase fraction under pressure correlates with disappearance of the superconductivity in FeSe$_{0.5}$Te$_{0.5}$ above 8.0 GPa. Curiously, heavy doping with Te does not essentially affect a structural transformation from the tetra- to hexagonal type under pressure. Thus, Te doping cannot be considered equivalent to physical pressure, despite of the big difference in ionic radii of Se and Te. A drastic increase of $T_c$ observed up to 1.5 GPa agrees well with previously reported data [8]. In the pressure range between 1.5 GPa and 6.0 GPa $T_c$ is almost pressure independent, although the lattice parameters decrease monotonously with pressure. However, in contrast to a previous study [8], pressure above 6.0 GPa applied at room temperature causes a transformation into the non-superconducting hexagonal phase.

Remarkably, neither FeSe, nor FeSe$_{0.5}$Te$_{0.5}$ do exhibit long-range magnetic ordering at ambient or under applied pressure in contrast to related arsenide phases [27, 28]. On the other hand, different experimental studies indicate that the superconducting pairing mechanism in FeSe and similar systems is related to magnetic fluctuations. The latter dramatically enhance under pressure leading to a strong raise of $T_c$ in FeSe [29, 30]. In case of high quality single crystalline FeSe$_{0.5}$Te$_{0.5}$ reported here, $T_c$ increases rapidly as a function of pressure ($0 < p < 2.0$ GPa) up to about its maximal value (ca. 20 K) showing similarity to FeSe. In the pressure range 2.0 GPa $< p <$ 5.0 GPa $T_c$ remains almost constant up to its sudden drop due to the transition into the high-pressure hexagonal phase above 7.0 GPa.

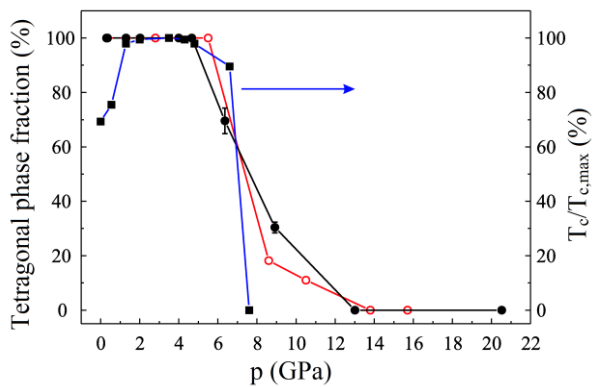

**Figure 8** Pressure dependence of tetragonal phase fraction obtained from Mössbauer measurements at 295 K for FeSe [2] (closed circles) and FeSe$_{0.5}$Te$_{0.5}$ (open circles) together with normalized $T_c$ for FeSe$_{0.5}$Te$_{0.5}$ obtained in magnetization measurements (closed squares).

**5 Conclusions** The pressure study of single crystalline FeSe$_{0.5}$Te$_{0.5}$ reveals distinct regions of $T_c(p)$ dependence presumably predetermined by miscellaneous structures. The rapid growth of $T_c$ in the range of small pressures observed by previous studies [8] could be explained by the enhancement of magnetic fluctuations by application of pressure [29, 30]. The subsequent plateau in $T_c(p)$ dependence is fairly broad and, comparing with conductivity measurements of powder samples under pressure, could be considered more plausible. In contrast to the known pressure studies [8], we found a pressure-induced structural transformation to the hexagonal NiAs-type polymorph phase which terminates superconductivity in FeSe$_{0.5}$Te$_{0.5}$. Prerequisite condition of the tetragonal to hexagonal transformation is pressure application at room temperature. The first-order reversible transition has the same features as non-substituted FeSe and shows a non-equivalence of physical pressure and chemical pressure due to Te doping.

**Acknowledgements** This work was supported by the Deutsche Forschungsgemeinschaft (DFG) through grants KS51/2-2 and ME3652/1-2 (priority programme SPP-1458). T.P. gratefully acknowledges the support from the Polish National Science Centre within the project No. 2012/05/E/ST3/02510 (programme "SONATA BIS").